\magnification=1200

\font\twelverm=cmr12 at 20pt

\def\rmi{\twelverm}

\input psfig
\vskip 3 pc
\centerline{\rmi The long-term optical behavior of MRK421}
\vskip 1 pc
\centerline{ F. K. Liu$^{1,2}$,  B.F. Liu$^{2,*}$,  and G.Z. Xie$^{2,3}$}
\vskip 6 pt
\centerline{1 International School for Advanced Studies, Via Beirut
2-4, 34013 Trieste, Italy;}
\centerline{ e-mail: fkliu@sissa.it}
\vskip 3pt 
\centerline{2 Yunnan Observatory, Academia Sinica, P.O.Box 110, 
Kunming 650011, China }
\vskip 3pt
\centerline{3 Center for Astrophysics, CCAST(World Laboratory), Beijing,
China}    
\vskip 3pt  
\centerline{* Max-Planck-Institut f\"ur Astrophysik, 
Karl-Schwarzschild-Str. 1, D-85740 Garching,}
\centerline{Germany; e-mail: lbf@mpa-garching.mpg.de}	

\

\

\centerline{\bf ABSTRACT}
{All data available in B band for the BL Lac object MRK421 from 22 
publications are used to construct a historical light curve, dating
back to 1900. It is found that the light curve is very complicated and
consists of a set of outbursts with very large duration. The 
brightness of MRK421 varies from 11.6 magnitude to more than 16
magnitude. Analyses with Jurkevich method of computing period of 
cyclic phenomena reveal in the light curve two
kinds of behaviors. The first one is non-periodic with rapid, violent
variations in intensity on time scales of hours to days. The second one
is periodic with a possible period of $23.1\pm 1.1$ years. Another possible
period of $15.3\pm 0.7$ years is not very significant. We have tested
the robustness of the Jurkevich method. The period of about one year
found in the light curves of MRK421 and of other objects is a spurious 
period due to the method and the observing window. We try to explain the period of $23.1
\pm1.1$ years under the thermal instability of a slim 
accretion disk around a massive black hole of mass of $2 *10^6 M_\odot$. 
}
\vskip 5 pc
\centerline{Accepted for publication in A\&A Supplement Series}
\vskip 8pc
\vfill
\par\break

\vskip 12 pt
\noindent{\bf 1. Introduction}
\vskip 6 pt
Accretion disks play a fundamental role in the theoretical models of
Active Galactic Nuclei (AGNs). Investigations indicate that
instability and pulsations in the inner, transonic regions with
parameters in proper ranges have various time scales (a review see
Wallinder et al. 1992). The time scale of the dwarf nova type limit 
cycle instability in AGNs is too long to be observed directly for a 
central black hole of mass of $10^6 M_\odot$ or higher
(Meyer-Hofmeister 1993), but the global thermal limit cycle 
(oscillation) time scale in slim accretion disks with $\alpha \sim 0.1$
around a central mass of  $10^6 M_\odot$ to $ 10^{10} M_\odot$ 
varies from a few years to several hundred thousand years (Honma et 
al. 1991). The latter periodic instability may be observed in some
QSOs and BL Lac objects with available observational data for around 
100 years. 

Rapid and large amplitude variability for BL Lac objects, a special
subclass of AGNs, has fostered considerable interest. The intra-day 
variability has been extensively investigated (see, e.g. Wagner and 
Witzel 1995 and references therein). Only a few investigations,
however, deal with the long-term variations of BL Lac objects due
to the lack of data available over a  long enough time scale.
Therefore, whether periodic or quasi-periodic fluctuations on long
term time scales exist is  unclear. Several years ago, we 
started collecting observational data on some BL Lac objects and 
investigated their long time-scale variability. Liu et al. (1995) 
showed that ON231 is a very active object with a probable periodic
activity of $13.6 \pm 1.3$ years.  This was interpreted
as a  thermal limit cycle (oscillation) in a slim accretion disk (for an 
alternative interpretation to the periodicity of OJ287, see 
Sillanp\"a\"a et al. 1988a). In this paper, we  show that the BL 
Lac object MRK421 is also very active and probably has periodic
activities. 

The X-ray selected BL Lac object MRK421, at $z=0.0308$ has attracted
much attention after it was identified as a BL Lac object. It is one
of the objects simultaneously observed at all electro-magnetic 
frequencies and one of the few objects radiating strong Gamma-rays.
 Many  observations to search for its 
variability in optical wave-band have been performed. Data from 
the archive plate 
collection of Harvard College Observatory showed a large range of 
variations,  $\Delta B \geq 4.7$ magnitude (Miller 1975). After 
examining the data between 1974 and 1982, Gagen-Torn et al. (1983) 
concluded that MRK421 varied in B band with an amplitude of 1.5 
magnitude on characteristic time scales of days to years. The 
purpose of present paper is to probe
the long time-scale variability of MRK421. We collect
all available observational data in B band and give a general 
discussion on the light curves in section 2. In section 3 we present
a detailed analysis of the light curve, through the Jurkevich ${V_m}^2$ 
test. In order to test the robustness of the Jurkevich method, we
analyze a simulated data set  in section 4.  
Final conclusions are given in section 5.

\vskip 2pc
\noindent{\bf 2. Variability analysis of long-term light curves}
\vskip 6pt
All observational data in B band on MRK421 are available in the following
studies: Miller (1975), Ulrich et al. (1975), Veron \& Veron (1975),
Veron \& Veron (1976), Miller et al. (1977), O'Dell  et al. (1978),
Mufson et al. (1980), Puschell $\&$ Stein (1980), Zekl et al. (1981), 
Gagen-Torn et al. (1983), Sitko et al. (1983), Cruz-Gonzalez $\&$ 
Huchra (1984), Moles et al. (1985), Sitko et al. (1985), Makino et
al. (1987), Xie et al. (1987), Sillanp\"a\"a et al. (1988), Xie et al.
(1988), Sillanp\"a\"a et al. (1991), Sitko $\&$ Sitko (1991), Takalo
(1991), and Takalo et al. (1992). Data for MRK421 consist of 565 
observations, dating back to 1900. Since we are searching for the long
time variability, we include those data estimated from Miller's
figures with relatively large date uncertainties (less than one month).
The B-band observations are used in this paper because there are more 
data available in B-band than in other bands . We 
translate the photographic magnitude $m_{pg}$ by the approximate relation
$m_B = m_{pg} +0.11 $ and the flux density, $f_B$, by $m_B = [\log 4.49 
\times 10^6 -\log f_B (mJy)]/0.4$ (Sitko, Schmidt, $\&$ Stein, 1985).

The error caused by the conversion from photographic to  photo-electric
values is not larger than 0.2 magnitude. The object does not produce a stellar 
image in deep photographic exposures, so  photometric data obtained 
with different entrance sizes are different. About half of our data are taken
from Miller's paper (Miller 1975), where the uncertainty of the iris 
photometry measures are typically 0.1--0.2 magnitudes. A large fraction of the 
remaining data are 
obtained with a diaphragm of about $26''$, within which the
contribution of the host galaxy is less than 0.2 magnitude. Thus the difference
between magnitudes derived through different entrance diaphragm sizes is 
less than
about 0.2 magnitude. Therefore, the photometric and photo-electric data
 are consistent within 0.2 magnitude, a very small value 
compared to the large range of variation of the object, 
$\Delta B \ge 4.7 $ mag. The magnitude uncertainties introduce 
noise and introduce an uncertainty on the parameters of the temporal features possibly detected on  the Jurkevich plot (see next section). 

The long-term light curve is shown in figure 1a. Because of our
purpose to investigate large-amplitude variations, we do not show individual
error-bars. The effect of errors on the periodicity analysis 
 will be discussed in section 3.  MRK421 is
very active, with a range of variation of $\Delta B \geq 4.7$ mag. The 
source reached a  maximal brightness of 11.6 mag in 1934 January and
was brighter than 12.5 mag on three occasions in 1901, 1916 and 1936
(Miller 1975).  After reaching
a maximum B=12.75 in 1982 April, MRK421 faded out until 1986 .
There are fewer  observations available for MRK421 after 1986, in B
band. The observations, however, still show that the source brightened
again (Takolo,1991, \& Takolo et al. 1992). 

To reduce small amplitude intra-day's fluctuations , we
averaged the light curve over 1 day. No significant difference has
been found. In order to probe the long-term 
behavior of the variations, we averaged the light curve over 100-day 
( Fig. 1b).  Because of the different 
quality of our data at different epochs, the impact of flickering in recent
data is washed out,while still significant in the early data. As the
object varies in intensity by about 0.5 mag on a time scale of several
hours (Xie et al 1988),  the largest difference between early 
epoch data point and the mean value can be estimated only within an 
uncertainty of  0.5 mag.
We have averaged the light curve . The 
resulting light curves  are similar. The peaks in 1934, 1975 and 
1982 remain unchanged. The difference between  Figs. 1a and 1b is 
quite significant. It indicates that MRK421 suffers large intensity
variations  on a time scale of a few months. The variability of MRK421
shows  two modes: a short one with a time-scale of a few
months to several years and a longer one with a time scale of the order of
ten years. We will analyze the repetition of the bursts in the light
curve using the Jurkevich method (1971) in section 3. 

\vskip2pc
\noindent{\bf 3. Periodic analysis}
\vskip 6pt
 Is there any 
period in the light curve? In this section, we try to answer this 
question using the powerful Jurkevich $V_m^2$ method (Jurkevich 1971).

The Jurkevich method is based on the expected mean square deviation. It 
tests a run of trial periods around which the data are folded. All 
data are assigned to $m$ groups according to their phases around each
trial period. The variance $V_i^2$ of each group and the sum 
$V_m^2$ of all groups are computed. For a trial period equal to the 
true one, if any, $V_m^2$ reaches its minimum, and a ``good'' period 
will give a much reduced variance relative to those given by other 
false trial periods and with almost constant values. No firm rule 
exists for assessing the significance of a minimum in the $V_m^2$ plot. 

As in Kidger et al. (1992) and Liu et al (1995), we consider the parameter
$f$ ,
$$  
f = {(1-V_m^2) \over V_m^2} \eqno(1)
$$ 
where $V_m^2$ is the normalized value. In the normalized plot, a value
of $V_m^2=1.0$ implies that $f=0$ and hence there is no periodicity at
all. The best periods can be identified from the plot. A value $f \ge
0.5$ generally indicates that a strong periodicity exists in the data, 
whilst $f < 0.25$ usually indicates that the periodicity, if genuine, is a 
weak one. A further test is the relationship between the depth of the 
minimum and the noise in the ``flat'' section of the $V_m^2$ curve 
close to the adopted period. If the absolute value of the relative 
change of the minimum to the ``flat'' section is larger than ten times
the standard error of this ``flat'' section, the periodicity in the 
data can be considered as significant and the minimum as highly 
reliable. In the Jurkevich test the parameter m can be modified: more 
groups give higher sensitivity, but fewer data points per group introduce
a larger noise in the plot. So we analyze the data sample mainly
using m=10, which gives us over 50 points per group. To search for
short time scale periods, we choose a small interval between two
successive trial periods.

The result of the analysis with $m=10$ is shown in Fig. 2. A minimum
of $V_m^2 =0.653 $ ($f=0.532$) is significant at a trial period of 
$23.1\pm 1.1$ years. A similar analysis with $m=20$ shows that $V_m^2 
= 0.615$ ($f=0.627$) at the period of 23.5-year. In addition to the 
period of $23.1\pm 1.1$ years, the broad minimum at $P = 15.5$ years is
also significant with $V_m^2 = 0.701$ and $f=0.427$ but not as certain as
the one obtained at $ P = 23.1$ years. We have considered the half width at
half minimum as the ``formal'' error (c.f. Jurkevich 1971) to derive all
effects on the precision,
including random variations in the exact interval between outbursts,
poor coverage of some of the early outbursts and the larger error in some of
the early photographic photometry,  the uncertainty of observed data
 estimated from
the figures in the literatures, random variations in intensity, and the
changing width of the outburst structure.
The errors caused by the conversion from photographic to photo-electric
values and by the measurement with different diaphragms are considered in
this analysis  as
random variations in the intensity. They would reduce the depth of
minima and therefore the significance of the periodicity found . 
These errors also increase the ``formal'' error, and this effect has
 been taken into account. The fluctuations 
seen around the minimum may also be caused by flickering, which is definitely
non-periodic. The broad width of the minima  may also result 
from the broad structures of the bursts, the drift of the real period and the
effect of adjacent periods, if present.

To compensate for the heavy weighting of recent data, we use the 100-day
averaged light curve. This interval is long enough compared to the 
possible periods of 15.3 years and 22.8 years and unlikely to prevent a 
distribution of  
the long term variation findings. The  result of the Jurkevich test shows
a larger noise due to  fewer points in every group and
flickering effects in the early epochs. However, a  minimum of $V_m^2$ 
near the possible period of 22.8 years is still seen.

Considering the redshift of 0.0308, the period of 23.1 years corresponds to
22.4 years in the rest frame of the source. 

\vskip 2pc
\noindent{\bf 4. The robustness of the Jurkevich method }
\vskip6pt

In order to test the robustness of the Jurkevich method and to
investigate intermediate  time-scale periods, we exclude Miller's data and
use the observational data from 1972 to 1991 . 
During this period, MRK421 was more extensively monitored and thus has
sufficient data  for our analysis to give reliable results. 
 The result of the analysis, with $m=5$,
is shown on Fig. 3.

Although the time interval considered in this case is less than that covered in
Fig. 2 ( only  19 years ), a minimum at $P= 13.7 \pm 2.0 $ 
years is
quite significant and broad, and consistent with the results given in
Fig. 2. This  confirms the period of $15.3$ years and test the 
robustness of the Jurkevich method. 

In addition to the minimum at $13.7$ years, a second minimum at period
 $P = 6.0$ years with $V_m^2 = 0.606$ and $f=0.649$ is found to be 
significant and
broad. Its relative depth, however, is only about 8 times the nearby noise . 
Although the period of $P
= 6.0$ years is about half the period  of $P = 13.7$ year, we cannot be 
sure of the reality of the former, as it does not appear in Fig.2.  

In addition to the possible periods of $22.8$, $15.3$ and $6.0$ years,
the plots also show minima at $P = 1.1$, 2.2 and 3.4 years in Fig. 2
 and 3 with relatively less significance. A one-year period was also
found in the light curves of  ON231 (Liu et al. 1995) and 3C120
(Jurkevich et al 1971). A one year period is doubtful as the
astronomical cycle is of one year. In order to check whether the period
is a spurious result of the Jurkevich method, we did following test.

To test the method, we take an object  with only random variations 
 with an amplitude of 4 magnitudes. To mimic real
observations, we make a further assumption that the object can been 
observed only from the beginning of January to the end of March every 
year and that the  available data
covers a one hundred year range. We also assume that, for moon light
 reasons for example, it can been observed only
for 10, 20 or 30 days a month. Under these assumptions, the number of
data points (one point a day) would be 3000, 6000 and 9000. The result
of the Jurkevich analysis for the 20-day case is shown in Fig. 4. The results
for the others are similar. No significant minima are found at one
year and multiple. When we change the assumption from three months to
four months and do the test again, the conclusion is unchanged. Now,
we assume that the source  varies sinusoidally  with a period 
of 12.5 years and we keep all the other assumptions . The result of
the analysis for the 20-day case is shown in Fig. 5. In addition to the 
minima at 12.5 years and multiple, the minima at one year and multiple
become very significant. If we assume we could observe 12 months a year, 
the minimum
corresponding to a period of one year does not exist any more in the 
$V_m^2$ plot. We conclude that the Jurkevich method 
does not give a spurious period of one year for a randomly variable 
source, but if there exists a long term period in the light curve
of the source, a spurious period of around one year will
appear. This can probably be understood as following: when the trial period is 
slightly different from half a year, and one year and its multiple, some
of the ten groups ($m$) contain very few observational points which 
lead to a very small variance $V_i^2$ and therefore small $V_m^2$ 
(cf. Fig. 5). So the minima at about one year and  multiple might
be taken as another signal of the existence of a long time-scale period 
in the light curve.

The Jurkevich analyses of the observational data with Miller's and 
without Miller's data provide similar results. These results are 
independent of the parameter $m$. Our  analysis shows
that probably two periods exist in the light curve of the BL Lac 
object MRK421: one of around 15 years and another of around 23 years.
However, the current data set covers only
four times  the possible period of 23 years, so
more data is needed to confirm this 23 years period. 

\vskip2pc
\noindent{\bf 5. Cnclusions}
\vskip 6pt

We assembled the historical light curve of the BL Lac object MRK421
and searched for its possible periodicity using the Jurkevich
method. Our results indicate that this object is very active and
probably has two periodic activities . One 
period is  of $23.1\pm 1.1$ years and the other is of $15.3
\pm 0.7$ years which, if real, superposes on the former. The former 
period has a higher confidence. The period of 23 years is about half 
the time interval between the well  observed outbursts in 1934 
and 1982. If the period is real, outbursts probably occurred 
between 1953 and 1968, where unfortunately, no published data are available. 
We must remember, however, that there is some noise on the curve
(Fig.2) and that the total observation range spans only 
four times  the period of 23 year. More 
observations are required to be assess the reality of this period.

The period of one year and multiple found in MRK421, in ON231 (Liu et al. 1995) and in
3C120 (Jurkevich et al 1971)  are spurious results  due to the 
existence of a long time-scale period and a cycle of 
one year in the astronomical optical observations. 

Regarding  
the 23 years period, we tentatively provide below a theoretical explanation. 
Sillanp\"a\"a et al. (1988a) suggest a binnary black hole model to explain
the quasi-periodic behaviour found in BL Lac object OJ287.  However, there
are several difficulties with the binnary model: observed periodicity is not
exact, the period in OJ287 corresponding to the minimium of brightness is quite
doubtful, observed burst structures are very broad, the system is short-lived
due to gravitational radiation and dynamically unstable due to the interaction 
between seconday black hole and disk.  Periodicity has probably been found in
many BL Lac objects (Liu 1996) and therefore isn't probably of the 
origin of binary black holes. The fact that the duration of a burst is around
half the quasi-period can be interpreted in terms of the thermal instabilities 
in a slim accretion disk in AGNs. Some simulations have shown that slim disks
can indeed be subject to limit-cycle type oscillations, as in the case of dwarf
novae although with a different oscillation behavior (Taam \& Lin 1984; Lasota
\& Pelat 1991; Honma et al. 1991). The basic characteristics
of the thermal limit cycles depend strongly on the viscosity parameter
$\alpha$, central black hole mass $M_6 = {M \over 10^6 M_\odot}$, 
accretion rate $\dot M$ and generalized stress tensor parameter $\mu$
(c.f. Wallinder et al. 1992). However, the time duration of the bursts
is almost independent of both $\mu$ and $\dot M$, and may be written 
empirically as 
$$
t_{burst} \simeq 4.5 \alpha_{0.1}^{-0.62} M_6^{1.37} \quad {\rm yrs},
\eqno(2) 
$$
when $\mu =0.5$ and $\dot M \simeq 0.2 \dot M_c$ where $\dot M_c = \dot
M_E /\epsilon$, $\dot M_E$ being the Eddington accretion rate and
$\epsilon$  the accretion efficiency (Honma et al. 1991). 
The time interval between 
subsequent bursts depends strongly on $\mu$, but weakly on $\dot M$.
As both the origin and the properties of the presumed viscosity in 
accretion disks are unknown at present, its hydro-magnetic origin is 
one of the options. Horiuchi and Kato (1990) suggest that $\mu \simeq
0.5$ may hold if the escape rate of the magnetic field is low. With 
these values of the parameters, the thermal limit cycle time $t_{cyc}$
(period) should be of the order of 2$t_{burst}$, i.e.
$$
t_{cyc} \sim 9.0 \alpha_{0.1}^{-0.62} M_6^{1.37} \quad {\rm yrs} . 
\eqno(3)
$$
For MRK421, if we adopt the typical values of $\alpha = 0.1$, $\mu =0.5$ 
and $\dot M \simeq 0.2 \dot M_c$ and search the central black hole
mass $M$ to get a period of 22.4 years, we find an estimated mass of 
$M \simeq 2 * 10^6 M_\odot$. This mass is reasonable if the parent
galaxies of BL Lac objects are FR I radio galaxies. 

\vskip 12pt
\hskip -20pt{\bf Acknowledgments}
\vskip 6 pt
{We are grateful to the referee, Dr. D. Alloin,  for her helpful comments.} 

\vskip12pt
\hskip -20pt {\bf References}
\vskip 6pt

\noindent Cruz-Gonzalez I., \& Huchra J.P., 1984, AJ,  89, 441

\noindent Gagen-Torn V.A., Marchenko S.G., Smekhacheva R.I. \&
Yakovleva V.A., 1983, Astro-

physics,  19, 111

\noindent Honma F., Matsumoto R., \& Kato S., 1991, PASJ 43, 147

\noindent Horiuchi T., \& Kato S., 1990, PASJ, 42, 661

\noindent Jurkevich I., 1971,  Ap\&SS,  13, 154

\noindent Kidger M.R., Takalo L., \& Sillanpaa A.,1992,  A\&A,  264, 32

\noindent Lasota, J.P., \& Pelat D., 1991, A\&A, 249, 574

\noindent Liu F.K. 1996, in Proceedings of the 21st Century Chinese Astronomy 
Conference, held in 

Hong Kong on 1-4 August, Cheng K.S. et al. (eds.), in press 

\noindent Liu F.K, Xie G.Z., \& Bai J.M., 1995,  A\&A,  295, 1

\noindent Makino F., Tanako Y., Matsuoka M., et al., 1987,  ApJ,  313, 662

\noindent Meyer-Hofmeister E., 1993, in Proceedings of 4th 
MPG-CAS Workshop on High Energy 

Astrophysics and Cosmology, B\"orner G.,
Buchert T. (eds.), MPA/P8, P.126

\noindent Miller H.R. 1975,  ApJ,  201, L109

\noindent Miller H.R., McGimsey B.Q., \& Williamon R.M., 1977,  ApJ,  217, 382

\noindent Mufson S.L., Wisniewski W.Z., Wood K., et al., 1980,  ApJ,  241, 74

\noindent Moles M., Garcia-Pelayo J., Masegosa J., \& Aparicio, A., 1985,  ApJS, 
 58, 255 

\noindent O'Dell S.L., Puschell J.J., Stein W. A., \& Warner J.W., 1978,  ApJS, 
38, 267

\noindent Puschell J.J., \& Stein W.A., 1980,  ApJ,  237, 331 

\noindent Sillanp\"a\"a A., Haarala S., \& Korhonen T., 1988,  A\&AS,  72, 347

\noindent Sillanp\"a\"a A., Haarala S., Valtonen M., Sundelius B., Byrd G.G.,
1988a, ApJ 325, 628

\noindent Sillanp\"a\"a A., Mikkola S., \& Valtaoja, L., 1991,  A\&AS,  88, 225 

\noindent Sitko M.L., Schmidt G.D.,\& Stein W.A., 1985,  ApJS,  59, 323

\noindent Sitko M.L., Stein W.A., Zhang Y.-X., \& Wisniewski W.Z., 1983,  PASP,  95, 724

\noindent Sitko M.L. \& Sitko A.K., 1991,  PASP,  103, No.660, 160

\noindent Taam R.E., \& Lin, D.N.C., 1984, ApJ, 287, 761

\noindent Takalo L.O., 1991,  A\&A,  90, 161

\noindent Takalo L.O., Sillanp\"a\"a A., \& Nilsson K., et al., 1992,  A\&AS, 94, 37

\noindent Ulrich M.-H., Kinman T.D., Lynds C.R., \& Rieke G.H., 1975,  
ApJ,  198, 261

\noindent Veron P., \& Veron, M.P., 1975,  A\&A,  39, 281 

\noindent Veron P., \& Veron, M.P., 1976,  A\&AS,  25, 287 

\noindent Wallinder F.H., Kato S, \& Abramowicz M.A., 1992,  A\&AR,
 4, 79 

\noindent Wagner S.J., \& Witzel A. 1995, ARA\&A, 33, 163

\noindent Xie G.Z., Li K.H., Bao M.X., et al., 1987,  A\&AS,  67, 17

\noindent Xie G.Z., LU R.W., Zhou Y., et al., 1988,  A\&AS,  72, 163

\noindent Zekl H., Klare G., \& Appenzeller I., 1981,  A\&A,  103, 342

\vfill
\par\break

\vbox {
\centerline{\hbox{
\hfill\psfig{figure=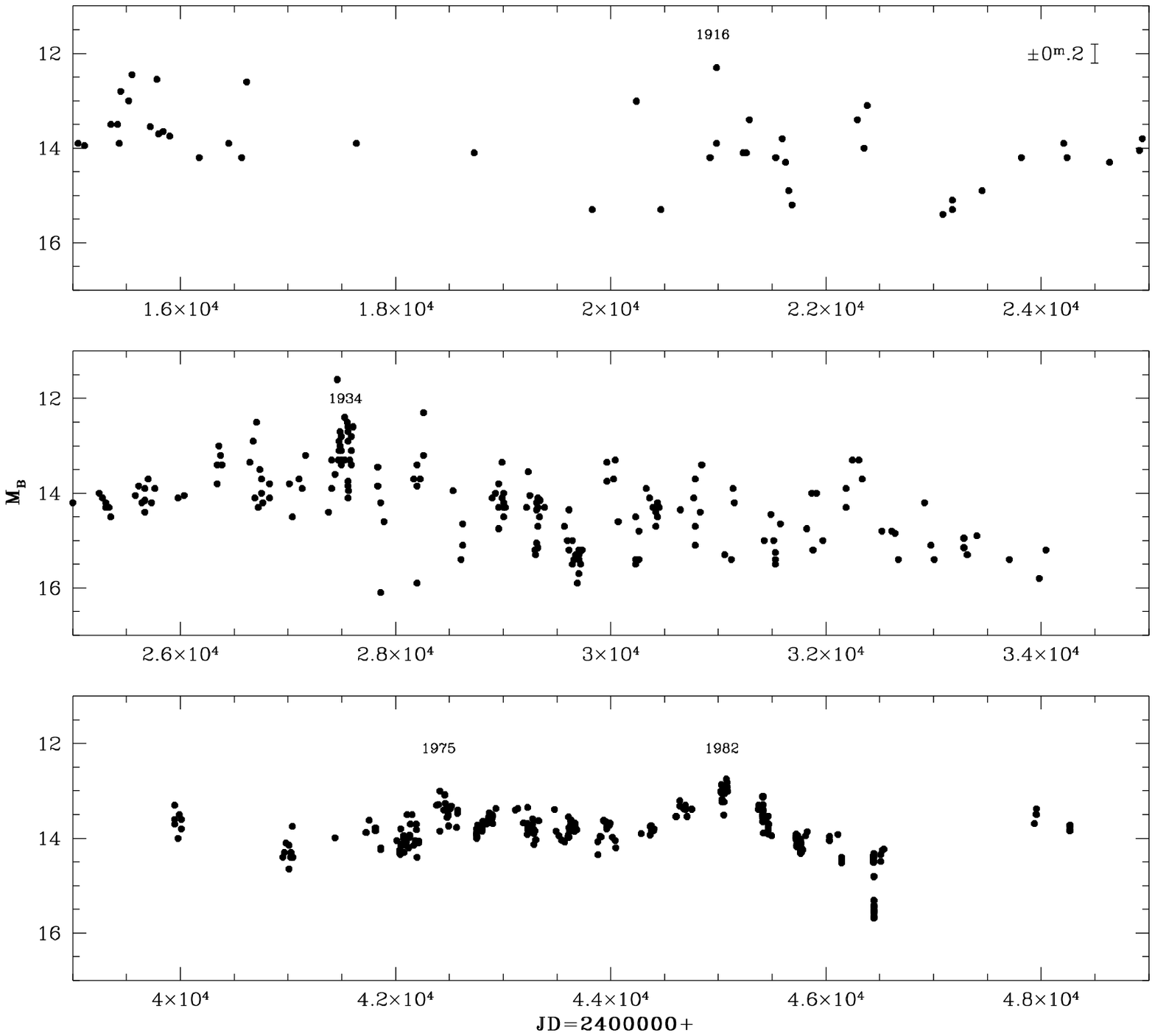,width=15.5cm}\hfill
}}
\vskip 3pc
{\bf Fig.1a} The long-term light curve of MRK421 from 1900
to 1991. The discontinuity of the light curve between 2435000 and
2439000 is due to lack of observations. }

\vfill
\par\break

\vbox {
\centerline{\hbox{
\hfill\psfig{figure=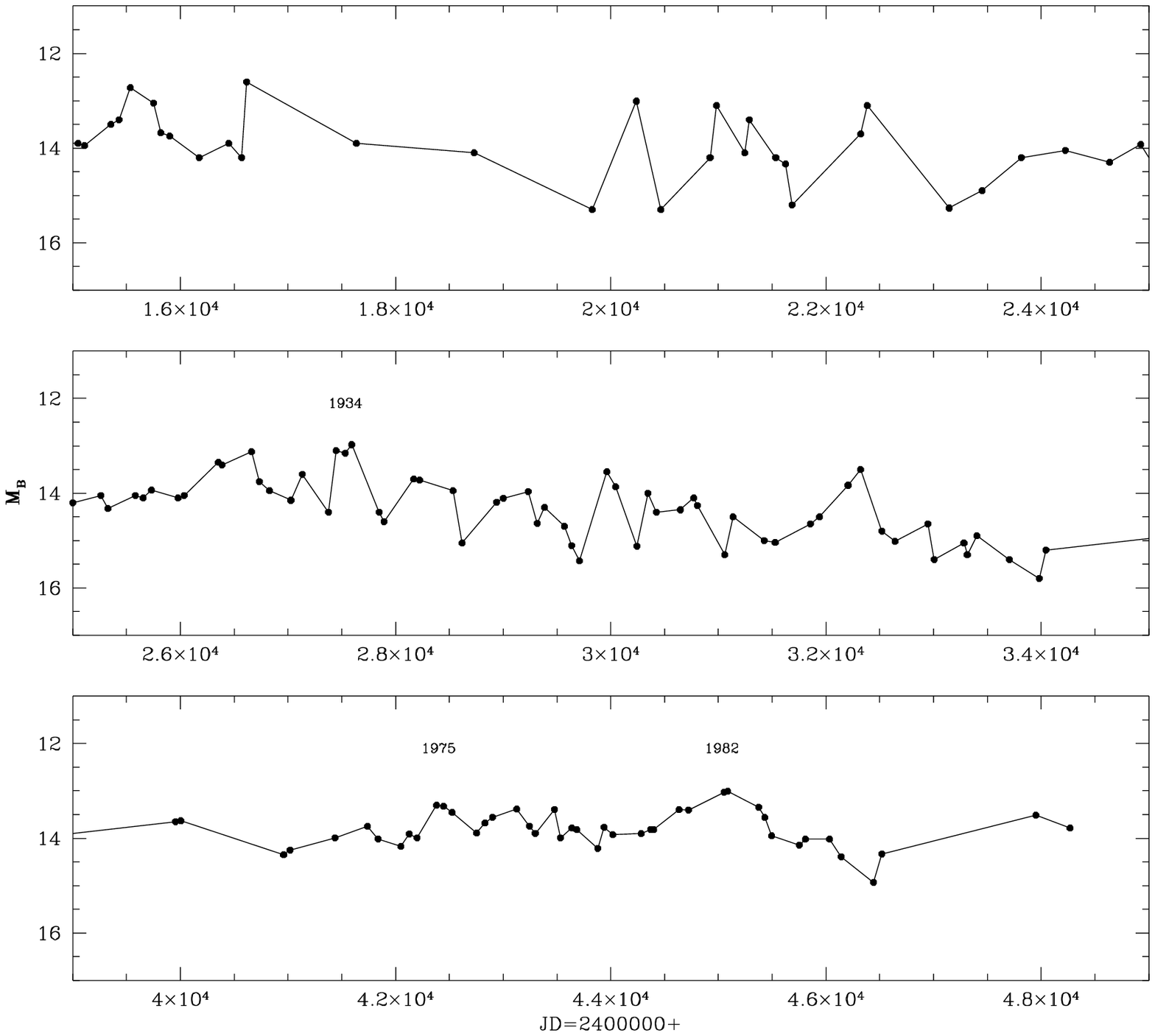,width=15.5cm} \hfill
}}
\vskip 3pc
{Fig.1b} The mean light curve of MRK421 over a 100-day mean.}

\vfill
\par\break

\vbox{
\centerline{\hbox{
\hfill\psfig{figure=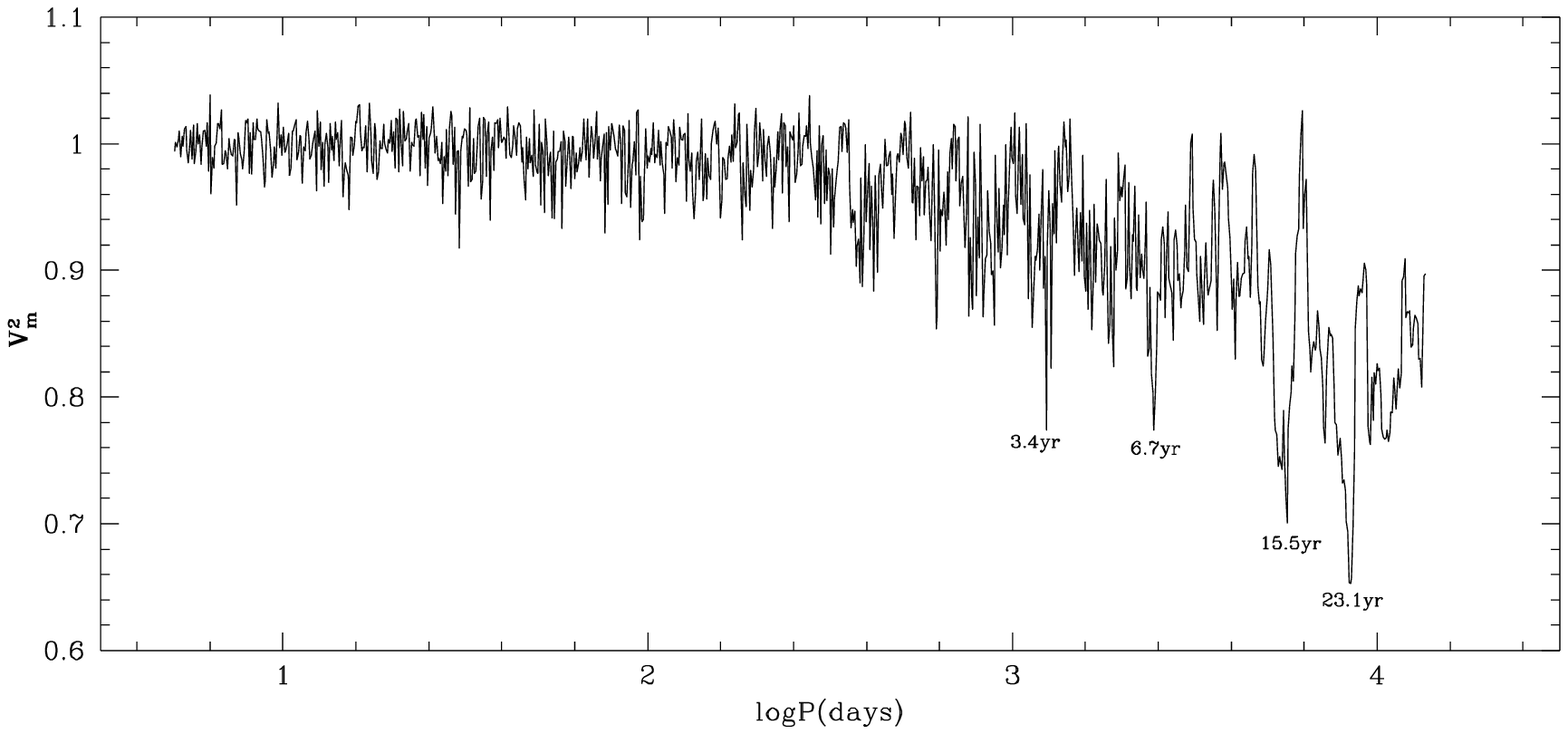,width=15.5cm}\hfill
}}
\vskip -7pc {\bf Fig.2} Results of the normalized Jurkevich test  for the 
period search, in  MRK421. The deepest minimum corresponds to a period 
of 23.1 year. The minima corresponding to periods of 3.4 years, 6.7 years, and 
15.5 years are also conspicuous.  }

\vfill
\par\break

\vbox {
\centerline{\hbox{
\hfill\psfig{figure=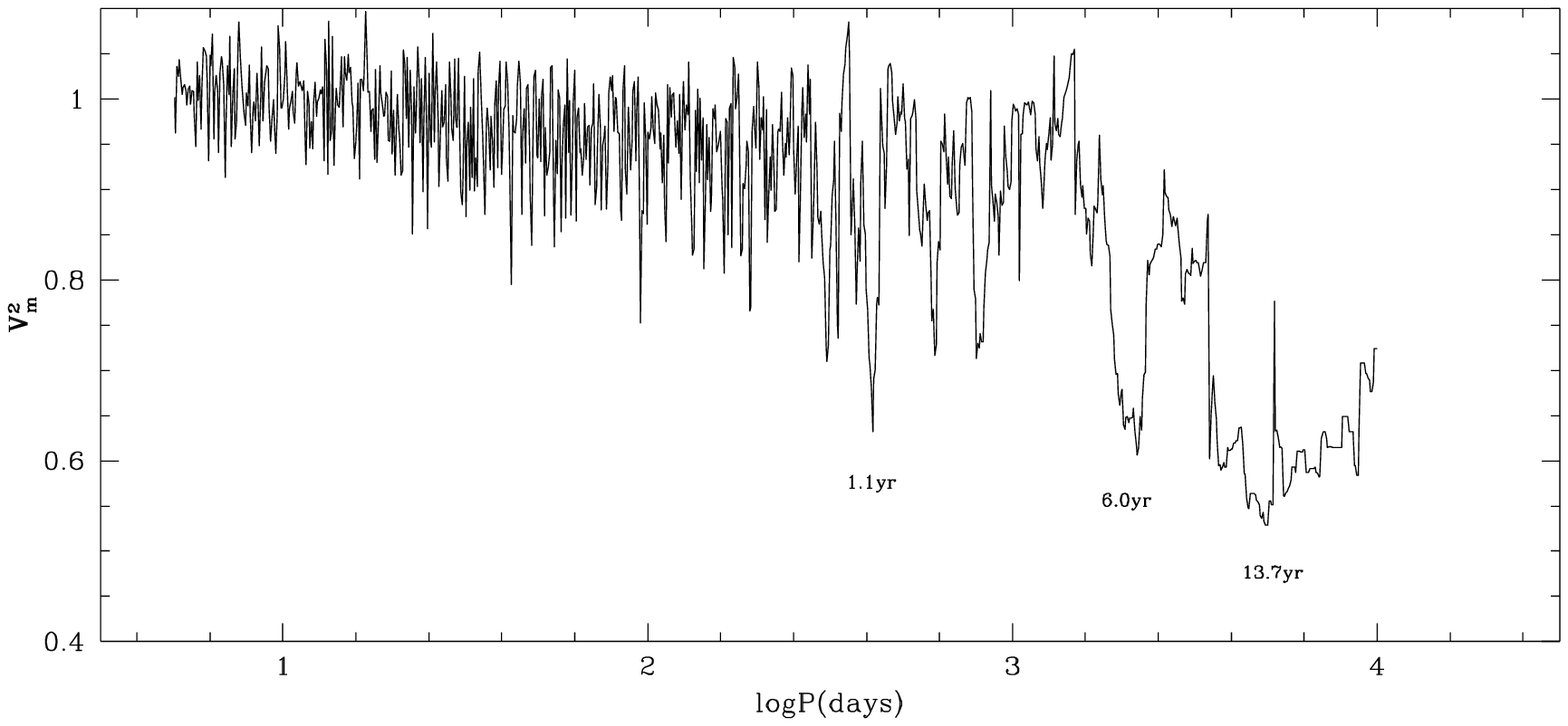,width=15.5cm}\hfill
}}
\vskip 4pc
{\bf Fig.3} Results of the normalized Jurkevich test  for the 
period search, in  MRK421, excluding Miller's data, with m=5. The  minima 
corresponding  to periods of 1.1, 6.0, and 13.7 years are significant.  }

\vfill
\par\break

\vbox {
\centerline{\hbox{
\hfill\psfig{figure=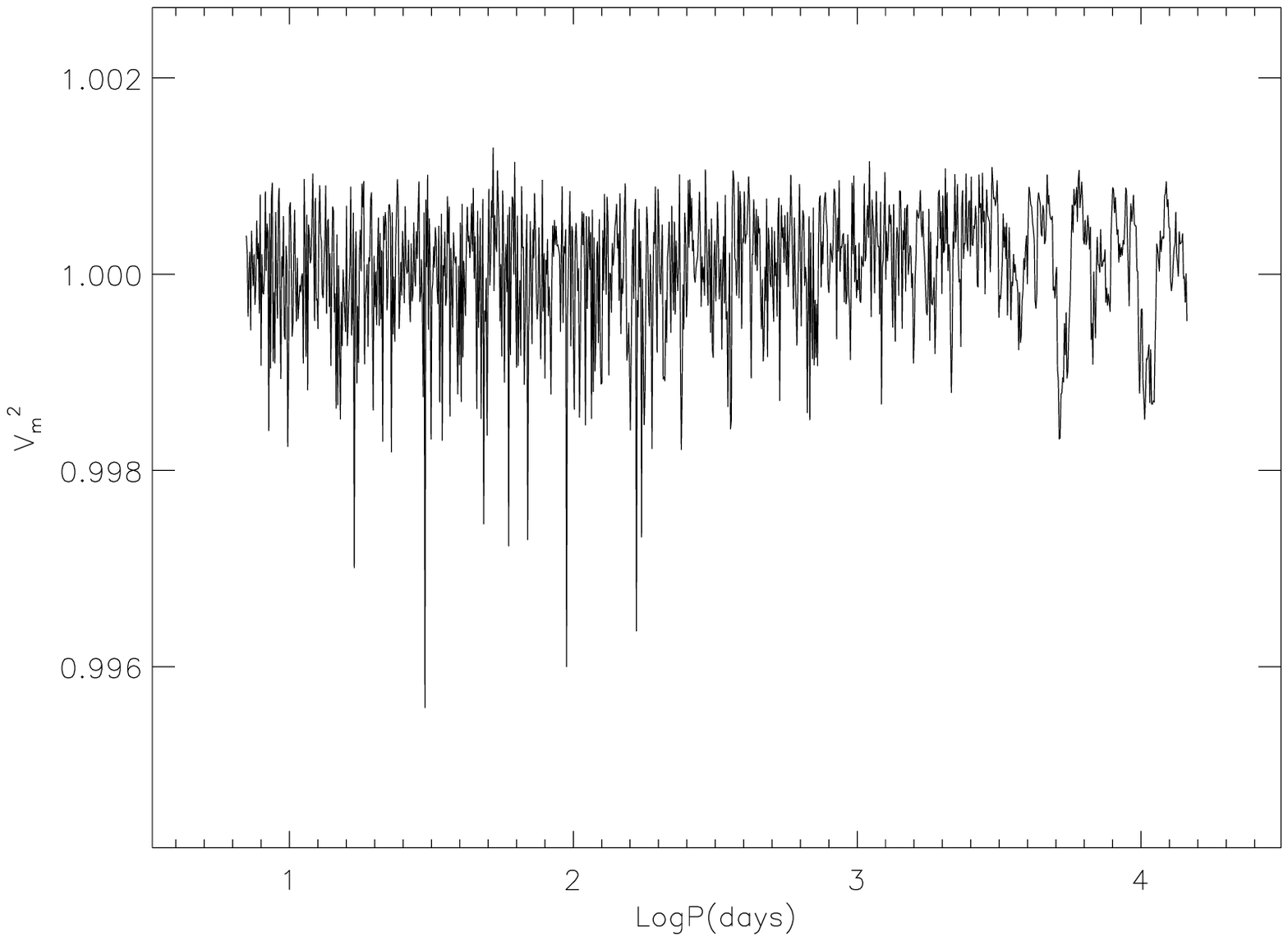,width=15.5cm}\hfill
}}
\vskip 6pc
{\bf Fig.4} The plot of normalized $V_m^2$ vs trial period
for a test object with  random variations . No
significant minima corresponding to trial periods of one year and 
multiple are found in the plot.    }

\vfill
\par\break

\vbox {
\centerline{\hbox{
\hfill\psfig{figure=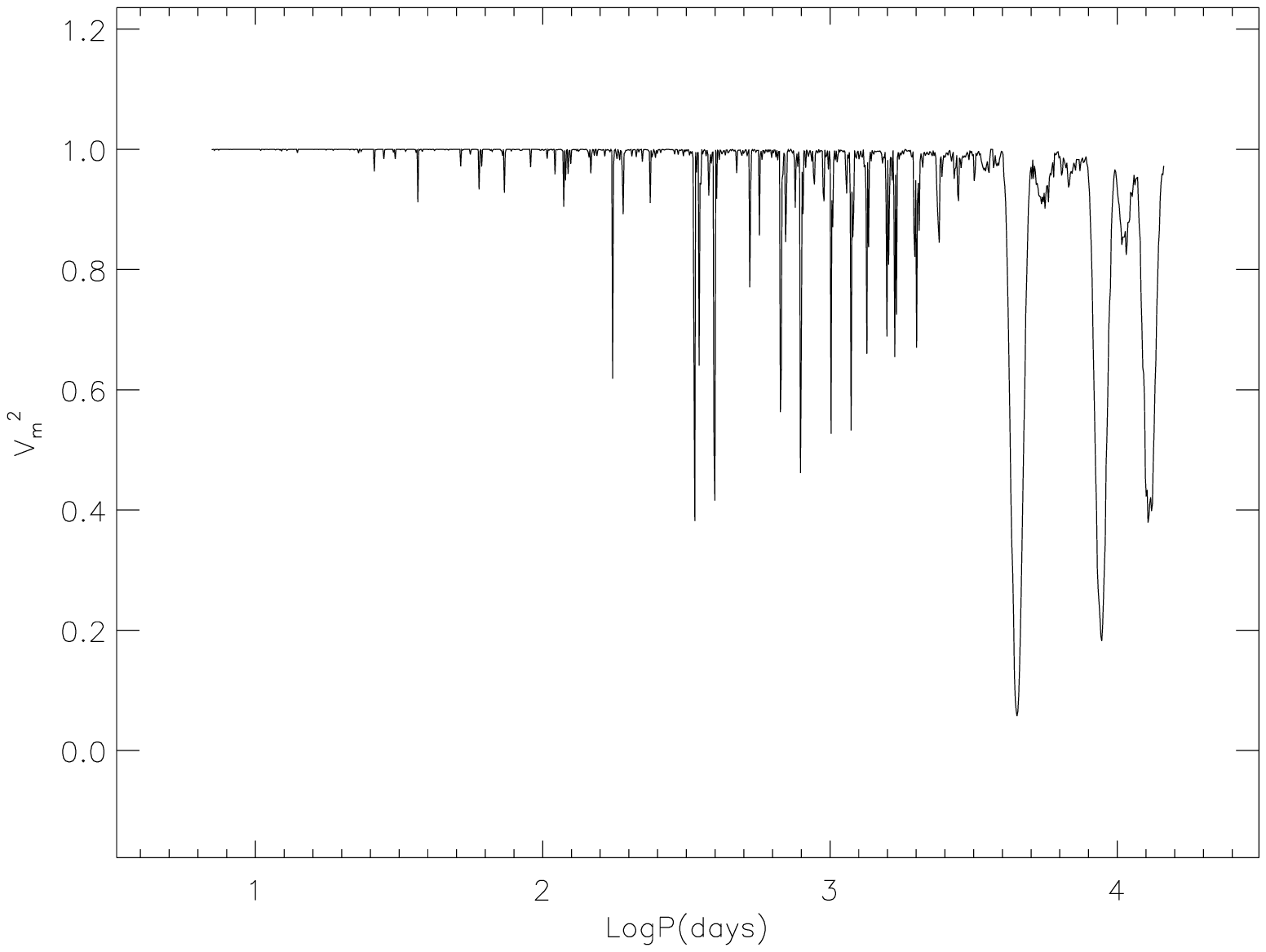,width=15.5cm }\hfill
}}
\vskip 6pc
{\bf Fig.5} Same as Fig. 4 for a test object with sinusoidal
variation of a period of 12.5 years. In addition to the minima at $P =
12.5$ years and multiple, the minima corresponding to a period of one
year and multiple are significant and are artifacts of the method
used. }

\bye